\documentclass[english,aps,pre,10pt,twocolumn,showpacs,preprintnumbers]{revtex4-1}
\usepackage{amssymb}
\usepackage[T1]{fontenc}
\usepackage{latexsym,epsfig}
\usepackage[latin9]{inputenc}
\usepackage{graphicx}
\usepackage{amsfonts}
\usepackage{epsfig}
\usepackage{dcolumn}
\usepackage{bm}
\usepackage{babel}
\usepackage{hyperref}
\usepackage{amsmath,mathrsfs}
\usepackage[usenames, dvipsnames]{color}

\makeatletter

\@ifundefined{textcolor}{}
{
 \definecolor{BLACK}{gray}{0}
 \definecolor{WHITE}{gray}{1}
 \definecolor{RED}{rgb}{1,0,0}
 \definecolor{GREEN}{rgb}{0,1,0}
 \definecolor{BLUE}{rgb}{0,0,1}
 \definecolor{CYAN}{cmyk}{1,0,0,0}
 \definecolor{MAGENTA}{cmyk}{0,1,0,0}
 \definecolor{YELLOW}{cmyk}{0,0,1,0}
 }

\makeatother

\begin{document}

\date{\today }
\title{Mobile-clogging transition in a Fermi-like model of counterflowing
particles}
\author{Roberto da Silva, Eduardo V. Stock}
\address{Institute of Physics, Federal University of Rio Grande do Sul,
Porto Alegre - RS, 91501-970, Brazil. }

\begin{abstract}
In this paper we propose a generalized model for the motion of a two-species
self-driven objects ranging from a scenario of a completely random
environment of particles of negligible excluded volume to a more
deterministic regime of rigid objects in an environment. Each cell of the
system has a maximum occupation level called $\sigma _{\max }$. Both species
move in opposite directions. The probability of any given particle to move
to a neighboring cell depends on the occupation of this cell according to a
Fermi-Dirac like distribution, considering a parameter $\alpha $ that
controls the system randomness. We show that for a certain $\alpha =\alpha
_{c}$ the system abruptly transits from a mobile scenario to a clogged state
which is characterized by condensates. We numerically describe the details
of this transition by coupled partial differential equations (PDE) and Monte
Carlo (MC) simulations that are in good agreement.
\end{abstract}

\maketitle

\section{Introduction}

\label{Section:Introduction}

The theoretical motion of particles in inhomogeneous media with local
impurities can be observed in many contexts in Physics and in a large number
of applications such as capture/decapture of electrons in micro,nano, and
meso devices \cite{Machlup1954,Kirton1989,noisesemiconductors}, the erratic,
but also directed, motion of molecules in chromatographic columns \cite%
{Cromatograph} and many others. However, other situations that consider
particles interaction can be explored, and we will particularly focus on the
ones that takes into account two different species of particles moving
against each other.

The patterns arising from counterflowing stream of particles can be studied
considering a wide range of apparently very different systems such as
pedestrian dynamics \cite{Pinho2016}, and charged colloids motion \cite%
{Vissers2011,VissersPRL2011}, which suggest more similarities between the
micro and macro systems that we can anticipate in this kind of modelling.
For this reason the straight formation of lanes, distillation, originated
from the complex emergent process of self-propelled and/or field-directed
objects/particles have raised a lot of interesting questions in the context
of statistical mechanics and the physics of stochastic process modelled by
Monte Carlo (MC) simulations or Partial Differential Equations (PDE) \cite%
{Stock2017,rdasilva2015}.

Similarly, more fundamental situations related to systems that collapse due
to clogging effects, related to the typical concentration of objects
phenomena, or their condensation patterns, under counterflowing streams,
lead to a fundamental question: How the environmental randomness compares to
the size of objects for the occurrence of clogging/jamming phenomena? In
order to understand this interesting problem that relates the micro with the
macro scales, in this work we consider a general modelling of streams of
counterflowing objects interpolating two very distinct situations:

\begin{enumerate}
\item \textbf{Situation (i)}: Objects of negligible sizes move in random
environment and their excluded volume is not a relevant parameter. In this
case the system is entirely random: the particle performs a biased random
walk in which it makes a step to the next cell with probability 1/2 or
remains at the original cell with the same probability. It works as if the
randomness of the motion is due to the resistance offered by the environment
and not from the interaction among objects. This randomness is then not
affected by the particle size.

\item \textbf{Situation (ii)}: The second extreme situation considers hard
objects where the excluded volume plays a role. It works as if rigid bodies
interact only with each other. In this situation the system is essentially
deterministic and one object does not move to a cell that does not have
enough space to allocate it.
\end{enumerate}

In this work we explore the transition between these two distinct scenarios,
by changing an external parameter denoted by $\alpha $. We are able to map
continuously from situation (i) ($\alpha =0$) to situation (ii) ($\alpha
\rightarrow \infty $) and the interesting point is how it is performed.

For that, we propose a simple model of two species of objects, denoted by $A$
and $B$, moving in opposite directions in a ring divided by cells, which
have the same maximum occupation denoted by $\sigma _{\max }$ that depends
on the stochasticity parameter $\alpha $. In order to describe $\alpha $,
which controls in which degree the maximal occupation $\sigma _{\max }$ can
be violated, we use an adapted Fermi-Dirac distribution that governs the
particle transition between the cells.

It is worth to emphasize that our model can represent a lot of different
systems, among them, for example, oppositely charged colloids in
counterflowing streams. In this context, the application of a strong
electric field along the longitudinal direction of the ring, would make
species $A$ drift, let us say, in a counter clock-wise fashion while species 
$B$ would drift in the opposite direction or we can picture objects entering
and leaving both extremities of a thin tape (tube) at a constant rate
(periodic boundary conditions), or as a last mention, the model proposed can
also mimic typical situations of pedestrians walking in subway corridors
under some peculiar conditions.

With our adapted Fermi-Dirac distribution we are able to describe the
complex dynamics of particles clogging in counterflowing that here is
studied in both: PDE and MC simulations. Our results indicate the existence
of a transition from a mobile phase (fluid) to clogged (condensate) phase,
by monitoring the evolution of the particle concentration patterns.

In section \ref{Sec:Model} we present the details of the model. Our main
results are presented in section \ref{Sec:Results}. Finally we summarize our
results in section \ref{Sec:Conclusions} and we present our main conclusions.

\section{The model}

\label{Sec:Model}

\begin{figure}[tbh]
\begin{center}
\includegraphics[width=1.0\columnwidth]{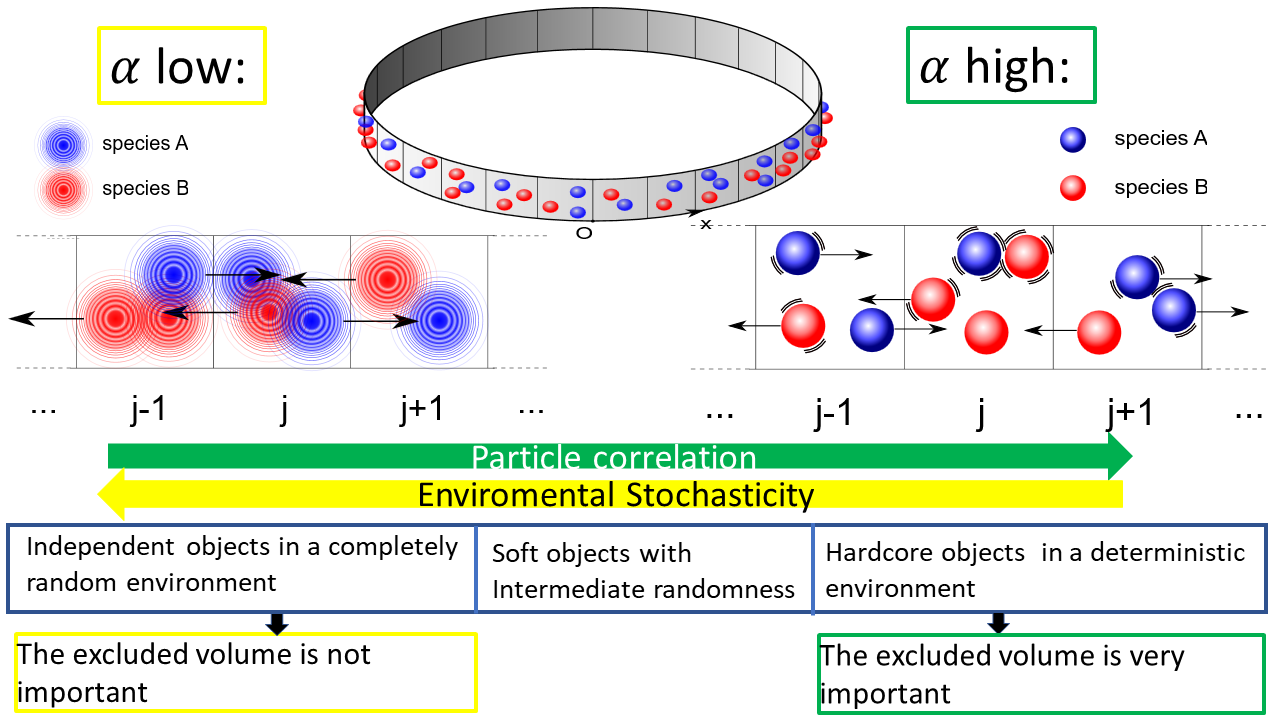}
\end{center}
\caption{Section of particles under counterflowing in a ring topology. The
two extreme regimes are illustrated: $\protect\alpha \rightarrow 0$, showing
that objects are independent of each other but interact with the stochastic
environment, and $\protect\alpha \rightarrow \infty $ where interacting
rigid bodies with high momenta ignore the randomness of the environment.
Intermediate values of $\protect\alpha $ correspond to some situation
between these two extremities. }
\label{Fig:all_plots}
\end{figure}

We consider in this paper a two-species model of particles which drift in
counterflow in an annular system composed by $L$ cells, each one with the
same limiting factor $\sigma _{\max }$, regarding the distinct situations
(i) and (ii) described in the previous section. For the sake of simplicity,
we pictorially illustrate our idea in Fig. \ref{Fig:all_plots}.

Considering that the concentration of particles (of whatever species) in the
following cell affects the locomotion of the particles in the current cell,
the concentration of target objects, according to our prescription can be
written by the recurrence relation: $%
A_{m,n}=p_{m-1,m}^{(n-1)}A_{m-1,n-1}+p_{m,m}^{(n-1)}A_{m,n-1}$, where $%
A_{j,k}$ is the density of particles of species $A$ of the cell $j$ at time $%
k$ and by construction $p_{m,m}^{(n-1)}+p_{m,m+1}^{(n-1)}=1$, since here $%
p_{i,j}^{(n)}$ denotes the probability of particle in cell $i$(position $%
x=i\varepsilon $) to transit to cell $j$ (position $x=j\varepsilon $) at $t=$
$n\tau $, where $\tau $ means the required time to perform such transition
and $\varepsilon $ is the step length.

Combining the equations, one has these equations%
\begin{equation*}
A_{m,n}-A_{m,n-1}=p_{m-1,m}^{(n-1)}A_{m-1,n-1}-p_{m,m+1}^{(n-1)}A_{m,n-1}
\end{equation*}

Taking into account that the occupation depends on a maximum level of
occupation, it is interesting to use the analogy of the Fermi level in the
context of conductor/semiconductor models, and that here it has a similar
role changing the temperature by a factor $\alpha ^{-1}$. If the desired
cell has a number of objects above the maximum level, the probability of
occupation behaves according to Fermi-Dirac occupation function: 
\begin{equation*}
p_{i,j}^{(n)}=(1+\exp [\alpha (\sigma _{j,n}-\sigma _{\max })])^{-1}\text{,}
\end{equation*}%
where $\sigma _{j,n}=A_{j,n}+B_{j,n}$ denotes the total number of objects at
the cell $j$, at the time $n$ which is the sum of the number of objects of
target species and the number of the objects of opposite species $B_{j,n}$.
The choice of Fermi-Dirac function to model the stochastic process of cell
occupation is very natural at this point: If the concentration of arrival
cell $A_{j,n}+B_{j,n}$ is greater than $\sigma _{\max }$, the transition is
hampered, otherwise the transition is facilitated. How much is hampered or
facilitated depends only on $\alpha $ which is not exactly the inverse of
temperature, but the matching between the Fermi-Dirac distribution and our
desired mapping is surprisingly meaningful. The objects can occupy the cell
even when $A_{j,n}+B_{j,n}>$ $\sigma _{\max }$. The only case this does not
occur is when $\alpha \rightarrow \infty $, since $p_{i,j}^{(n)}=1$ if $%
\sigma _{j,n}=A_{j,n}+B_{j,n}<\sigma _{\max }$, $p_{i,j}^{(n)}=1/2$ if $%
\sigma _{j,n}=\sigma _{\max }$, and $p_{i,j}^{(n)}=0$ when $\sigma
_{j,n}>\sigma _{\max }$. This case (corresponding to situation (ii)
previously considered), means that no more than $\sigma _{\max }+1$ objects
per cell are allowed. We consider that hard core objects interact and the
environment has no influence on their motion.

When $\alpha \rightarrow 0$, which corresponds to a low field regime, one
has $p_{i,j}^{(n)}=1/2$\ meaning that the objects do not interact with each
other, only with environment. In this case (corresponding to situation (i)
), we can imagine that an infinity number of objects (although not likely)
are allowed per cell, since the motion does not depend on $\sigma _{\max }$.
Here, the randomness in the environment. Since the interaction among the
particles is not considered in this limit, the objects are drifting in the
environment.

It is important to notice that our model prescribes other alternative
interpretations. One of them can imagine $\alpha $ as a kind of field that
drive the objects oppositely charged as considered for example in the
interesting stochastic lattice gas studied by Schmittmann, Wang, Zia \cite%
{Schmittmann1992}. In our modelling, we can imagine that for low $\alpha $,
the environment is important since the momenta of objects are low. On the
other hand when $\alpha $ is large the objects have high momenta and the
environment effects are not important to change the velocities of these
objects. In this case the interaction of objects have an important role.

So one has for the objects $A$, the recurrence relation $%
A_{m,n}=A_{m,n-1}+a_{m-1,n-1}-a_{m,n-1}$, and similarly for the objects $B$,
the relation $B_{m,n}=B_{m,n-1}+b_{m+1,n-1}-b_{m,n-1}$ with $a_{m,n}\equiv
A_{m,n}/\left[ 1+e^{\alpha(A_{m+1,n}+B_{m+1,n}-\sigma_{\max} )}\right] $ and 
$b_{m,n}\equiv B_{m,n}/[1+e^{\alpha(A_{m-1,n}+B_{m-1,n}-\sigma_{\max} )}]$.

Sure, by one can solve the recurrence relation as we do in this paper, but
we also analyze the corresponding differential equation. By considering the
situation $A_{m+1,n}+B_{m+1,n}\approx $ $A_{m-1,n}+B_{m-1,n}\approx
A_{m,n}+B_{m,n}$, we are led to a system of two coupled equations: 
\begin{equation}
\frac{\partial A(B)(x,t)}{\partial t}=-(+)C\frac{\partial }{\partial x}\left[
\frac{A(B)(x,t)}{1+e^{\alpha (A(x,t)+B(x,t)-\sigma )}}\right] \text{,}
\label{Eq:EDP}
\end{equation}%
where $C=\lim_{\tau ,\varepsilon \rightarrow 0}\frac{\varepsilon }{\tau }$.
It is important to notice that when $\alpha \rightarrow 0$, we have
uncoupled equations. In this situation, the solutions are expected to
satisfy $\frac{\partial A(x,t)}{\partial t}=-C\frac{\partial A(x,t)}{%
\partial x}$ and $\frac{\partial B(x,t)}{\partial t}=C\frac{\partial B(x,t)}{%
\partial x}$. For $C=1$, for example, under periodic boundary conditions, $\
A(x=L,t)=A(x=0,t)$ and $B(x=L,t)=B(x=0,t)$, it is easy to verify that the
only possibility is $A(x,t)=1$ and $B(x,t)=1$, with the initial conditions $%
A(x,t=0)=1$ and $B(x,t=0)=1$.

In the discrete formulation, this means $p_{i,j}^{(n)}=1/2$, which
corresponds to a ballistic behavior of objects, since if one considers $%
A(x,t=0)=L\delta _{x,0}$ and $B(x,t=0)=L\delta _{x,L}$, we expect the
solutions of recurrence relations, in the first round in the ring, to be
given by $A_{m,n}\approx L\binom{n-1}{m-1}2^{-n}$ and $B_{m,n}\approx L%
\binom{n-1}{L-m-1}2^{-n}$, once to execute $m$ success (It means to be in
the position $x=m\tau $) the particles have to execute $n>m$ experiments
according to a negative binomial \cite{Feller1966}. Here we are making the
number of cells exactly the length of tube (or simply $\varepsilon =\tau =1$%
).

However after several rounds, we must expect that $A_{m,n}$, $%
B_{m,n}\rightarrow 1$. But what happens when $\alpha $ increases and the
interactions between particles start to become important? We also expect
that $A_{m,n}$, $B_{m,n}\rightarrow 1$, which means that particles are
transiting without clogging in the channel?

In order to analyse this point we also consider performing Monte Carlo (MC)
simulations in order to support the numerical integration of the recurrence
equations. In these MC simulations, we consider synchronous updating (every
particle is verified if it goes to the next cell or stays stopped at the
same cell). In this case there is no possibility of the same particle to be
tested. The initial configuration considered in such simulations supposes
that all particles are randomly (uniformly) distributed among the cells.

Moreover, MC simulations make possible to analyze the clogging dynamics in
such kind of systems by considering an interesting order parameter that
measures a kind of current of objects over the annular tube, here called as
mobility, which is defined for $N$ particles at time $t$ as: 
\begin{equation}
M(t)=\frac{1}{N}\sum_{i=1}^{N}\xi _{i}(t)\text{,}  \label{Eq:mobility}
\end{equation}%
where $\xi _{i}(t)$ is a binary variable associated to particle $i$, that
assumes 0 if the particle stays stopped at time $t$ and 1 if this same
particle goes to the next cell at this same time. This quantity cannot be
calculated by solution of recurrent relations but when we perform MC
simulations it is easily obtained since we work exactly on the particles,
differently from recurrence relations. Some authors call this amount simply
as current.

First, let us to better explore the variation of $\alpha $. We solve the
recurrence relations starting with initial conditions $A_{m,0}=1$ and $%
B_{m,0}=1$, but with only one site $m=L/2$, empty, i.e., $%
A_{L/2,0}=B_{L/2,0}=0$. Differently from MC simulations we need an initial
defect to promote the time evolution of the system when we numerically
integrate the recurrence relations. So we expect that for low $\alpha $, $%
A_{m,n}\rightarrow 1$ (in our particular case, $A_{m,n}\rightarrow \frac{%
(L-1)}{L}\approx 1$, for $L$ large). On the other hand, i.e., for higher
values of $\alpha $, the question is: for which values of $\alpha $ the
system breaks down in a clogging situation?

In next section we present the main results of this work.

\section{Results}

\label{Sec:Results}

Let us initially consider the simplest case $\sigma _{\max }=1$, and let us
start by observing the density of particles $A$ and $B$ in both methods: MC
simulations and by numerical integrations of recurrence relations. A summary
of our main results can be seen in Fig. \ref%
{Fig:Different_frames_for_jamming}. Fig. \ref%
{Fig:Different_frames_for_jamming} (a) shows that for $\alpha =0.3$ the
system is freely flowing since for averaging both species over a large
number of runs ($N_{run}=100$) we have $A(x)\approx B(x)\approx 1$. It is
important to observe that for $N_{run}=1$ the fluctuations overcome the
expected behavior.

\begin{figure}[tbh]
\begin{center}
\includegraphics[width=1.0\columnwidth]{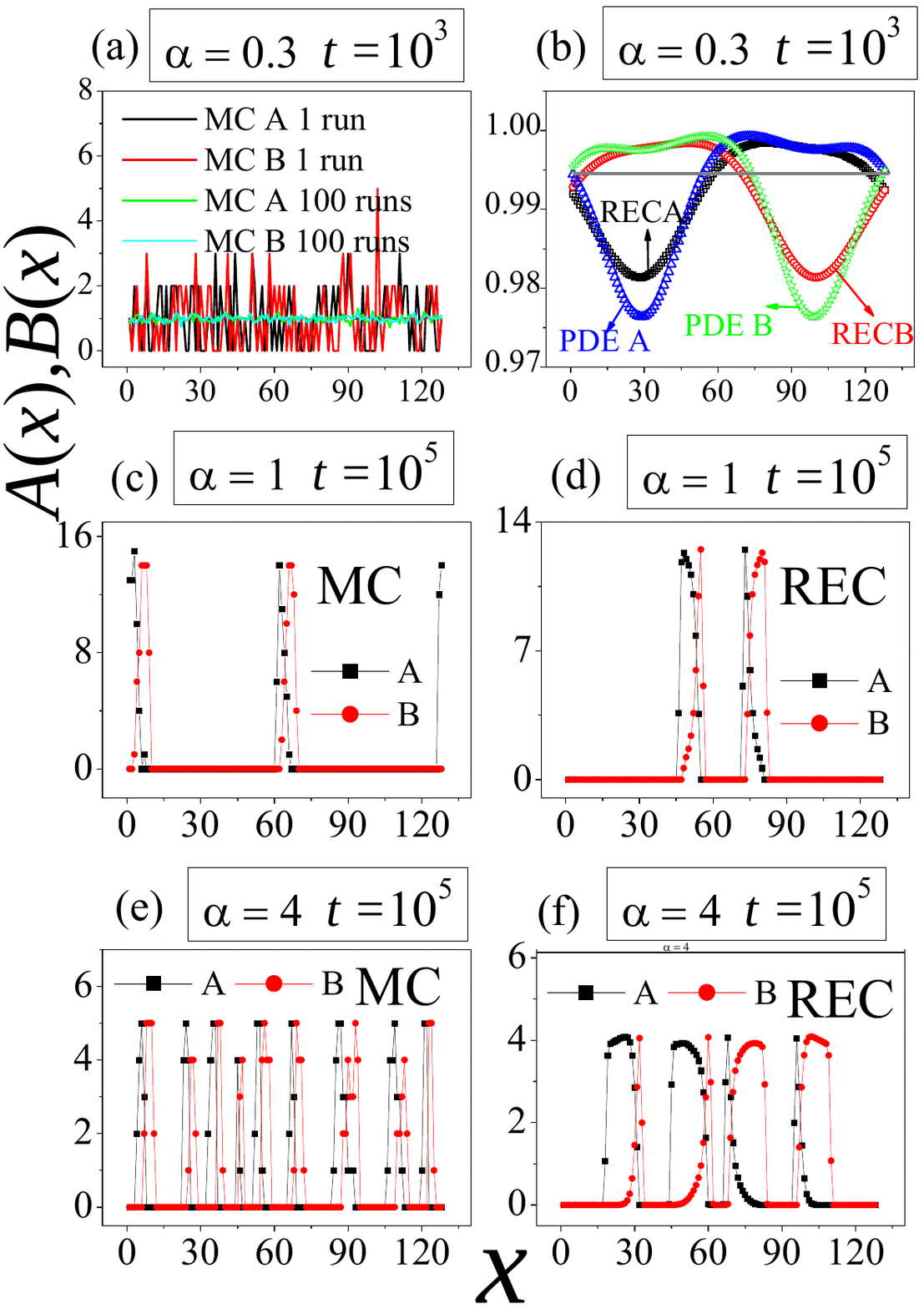}
\end{center}
\caption{Exploring the dynamics for different values of $\protect\alpha $,
methods and time steps. }
\label{Fig:Different_frames_for_jamming}
\end{figure}

The numerical solution was also obtained via two independent methods:
solution of PDE according to Eq. \ref{Eq:EDP} and numerical integration of
recurrence relations (REC) for the same value of $\alpha $, which have no
obligation to agree, but we expect that them to have at least the same
qualitative behavior. The results are shown in Fig \ref%
{Fig:Different_frames_for_jamming} (b). We can observe that although to
obtain the PDE we have changed the index in the recurrence relation, the
methods show curves around $A(x)=B(x)=1$ for intermediate time, $t=10^{3}$.
Moreover, in the same plot, for $t=10^{5}$ steps, the straight gray line,
represents all plots obtained from PDE and REC solutions that are
coincident, which indicates an exact agreement with $A(x)=B(x)=1$. From now
on, we will use only REC in this paper, since PDE only presents some
slightly differences in relation to the first one and it can be considered a
good representation of the model via partial differential equations. Other
mathematical properties of PDE in these counterflowing problems deserve
future exploration.

How about when $\alpha $ increases? For example for $\alpha =1$, both MC and
REC indicate some points of clogging characterized by high density of
particles $14\lesssim A\approx B\lesssim 16$ according to Figs. \ref%
{Fig:Different_frames_for_jamming} (c) and (d) showing that both methods
bring such situation. It is interesting, since when we have $\alpha =4$ the
jamming ocurrs with many situations of \textquotedblleft
bottlenecks\textquotedblright\ but now it occurs with lower intensity $%
3\lesssim A\approx B\lesssim 4$ exactly as shown in Figs. \ref%
{Fig:Different_frames_for_jamming} (e) and (f). So we raise the question,
about the existence of $0<\alpha _{c}<\infty $ for which the system transits
from mobile to clogged situation in the case $\sigma _{\max }=1$.

Here it is important to mention that differently from other works our
results look at transition dependence on the randomness of system ($\alpha $%
). Other works do not consider such parameter $\alpha $, and in pedestrian
dynamics the authors work with the transition of some parameter as average
system velocity or probability of clogging versus the density of pedestrian
(see for example interesting works: respectively \cite{Wei2015} and \cite%
{Marroquin2014})

\begin{figure}[tbh]
\begin{center}
\includegraphics[width=1.0%
\columnwidth]{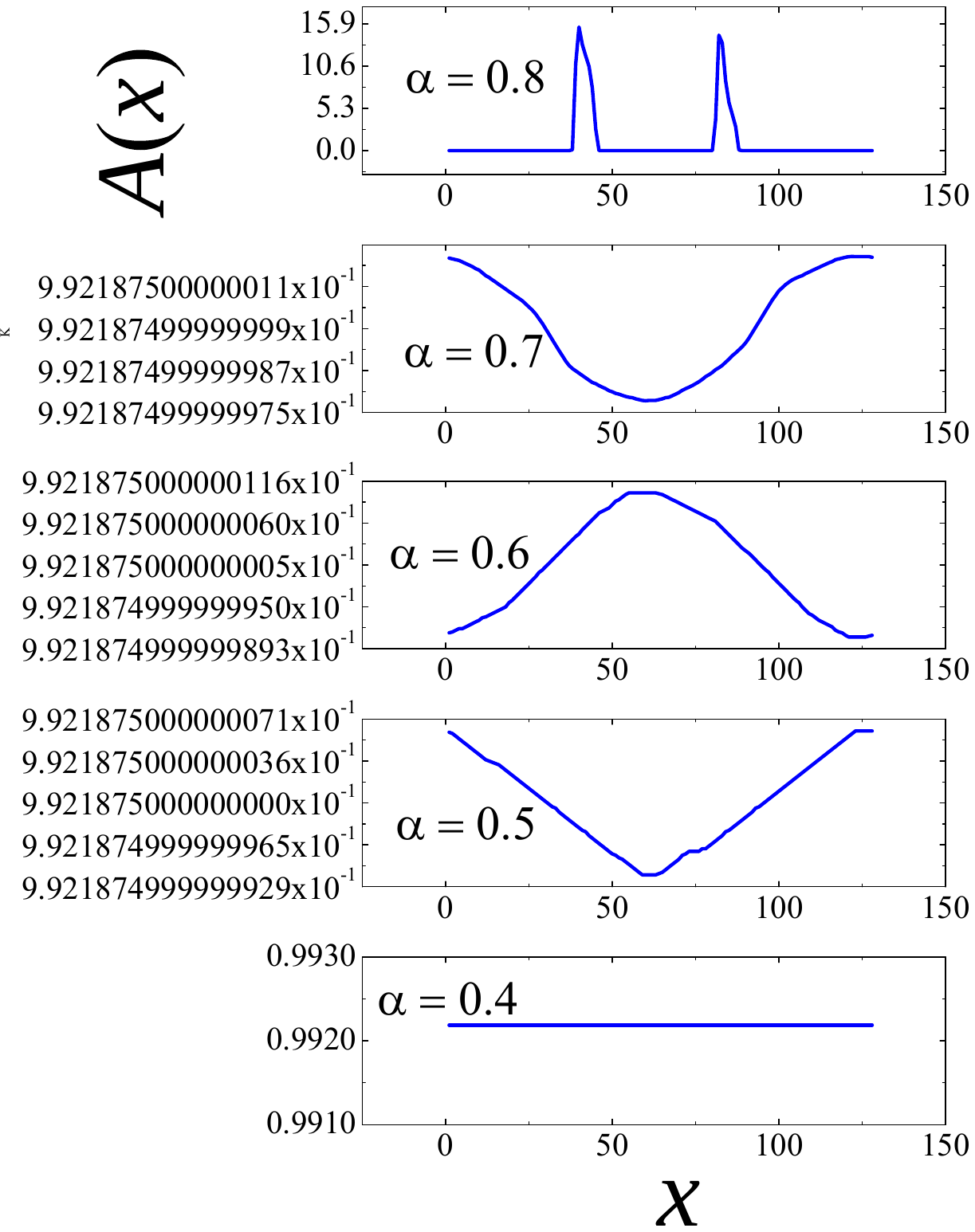}
\end{center}
\caption{Density of particles $A$ by REC solutions for $t=10^{5}$ iterations
(steady state). Clearly the systems is mobile for $\protect\alpha =0.4$ and
jammed for $\protect\alpha =0.8$ (formation of condensates). We can observe
a \textquotedblleft strange\textquotedblright\ behavior in the vicinity of
transition ($0.5\leq \protect\alpha \leq 0.7$). }
\label{Fig:Vicinity_of_transition}
\end{figure}

First by considering the vicinity of transition, we look the density of $A$
for five different values of $\alpha $ considering the stationary situation $%
t=10^{5}$ unit times. For example we have a mobile system for $\alpha =0.4$
and a system completely jammed for $\alpha =0.8$ (two pronounced peaks) but
for intermediate values of $\alpha $ (0.5, 0.6 and 0.7) the REC solutions
show that system seems to be in a metastable situation, where $A\approx 1$,
but slightly numerical differences are able to deform the solution leading
to strange shapes until to arrive the clogged situation starting from mobile
situation.

Thus, it goes into action, the mobility defined by Eq. \ref{Eq:mobility}. we
can discover what indeed is happening with the system. Here we look at time
evolution of the mobility considering a large number ($t_{\max }=10^{9}$ MC
steps) in order of our stop criterion to fail (what does not happen).
Basically the mobility arrives a steady state $M_{\infty }$ (stationary
mobility). We use a criterion to analyze when the system reaches this
stationary mobility. First, we observe the system visually, which
corresponds to a qualitative previous analysis. Secondly, we take a slope of
the stationary mobility by lags of 10$^{3}$MC steps. When the slope is
lesser than (in absolute vale) of $\eta $ we consider that system has
reached the stationarity. We use $\eta =10^{-7}$, and we check these cases
with our previous visual analysis. After these considerations we analyze the
behavior of $M(t)$, for fixed values of density as function of $\alpha $
taking the stationary value for each value.

\begin{figure*}[tbh]
\begin{center}
\includegraphics[width=1.0\columnwidth]{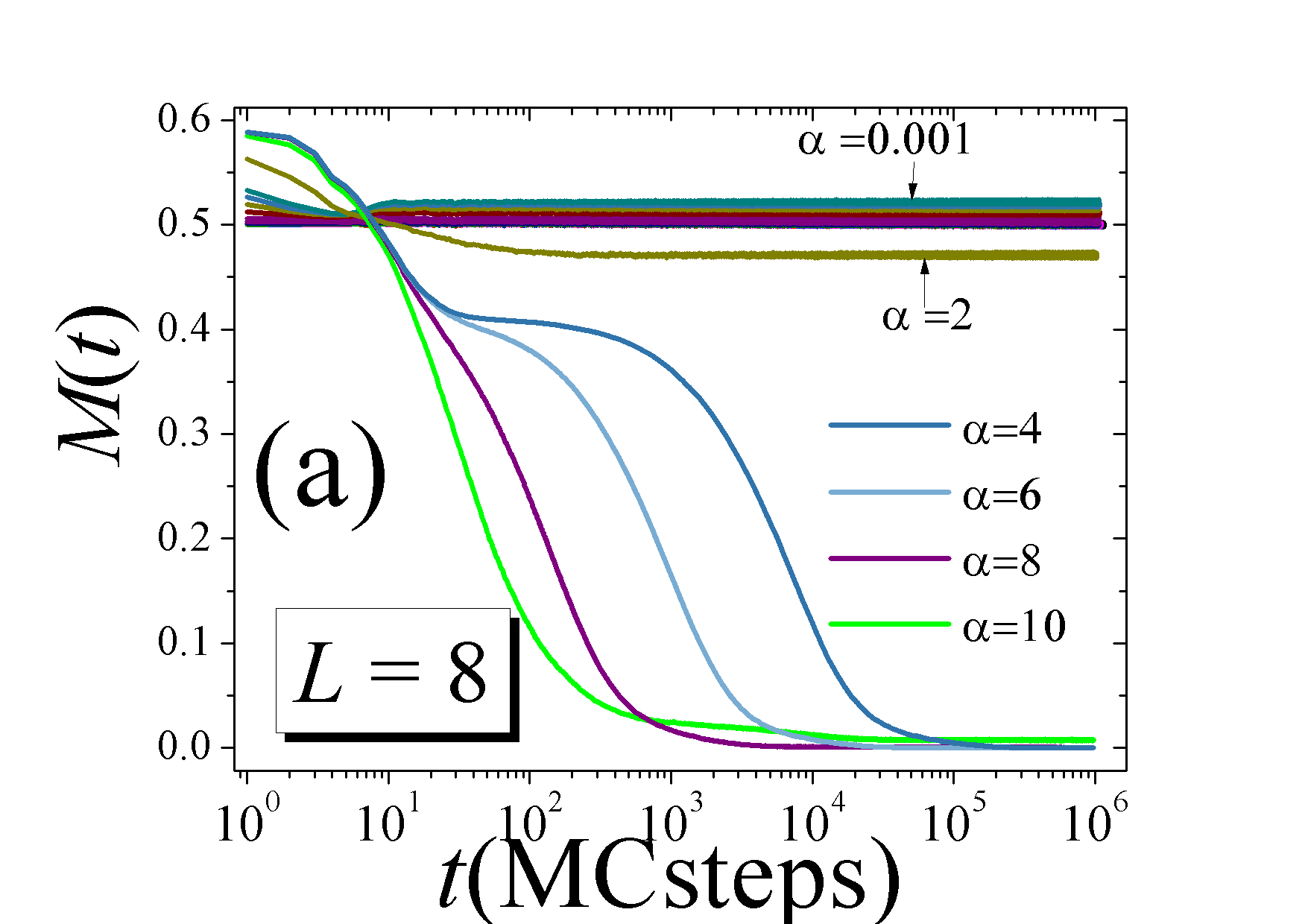}%
\includegraphics[width=1.0\columnwidth]{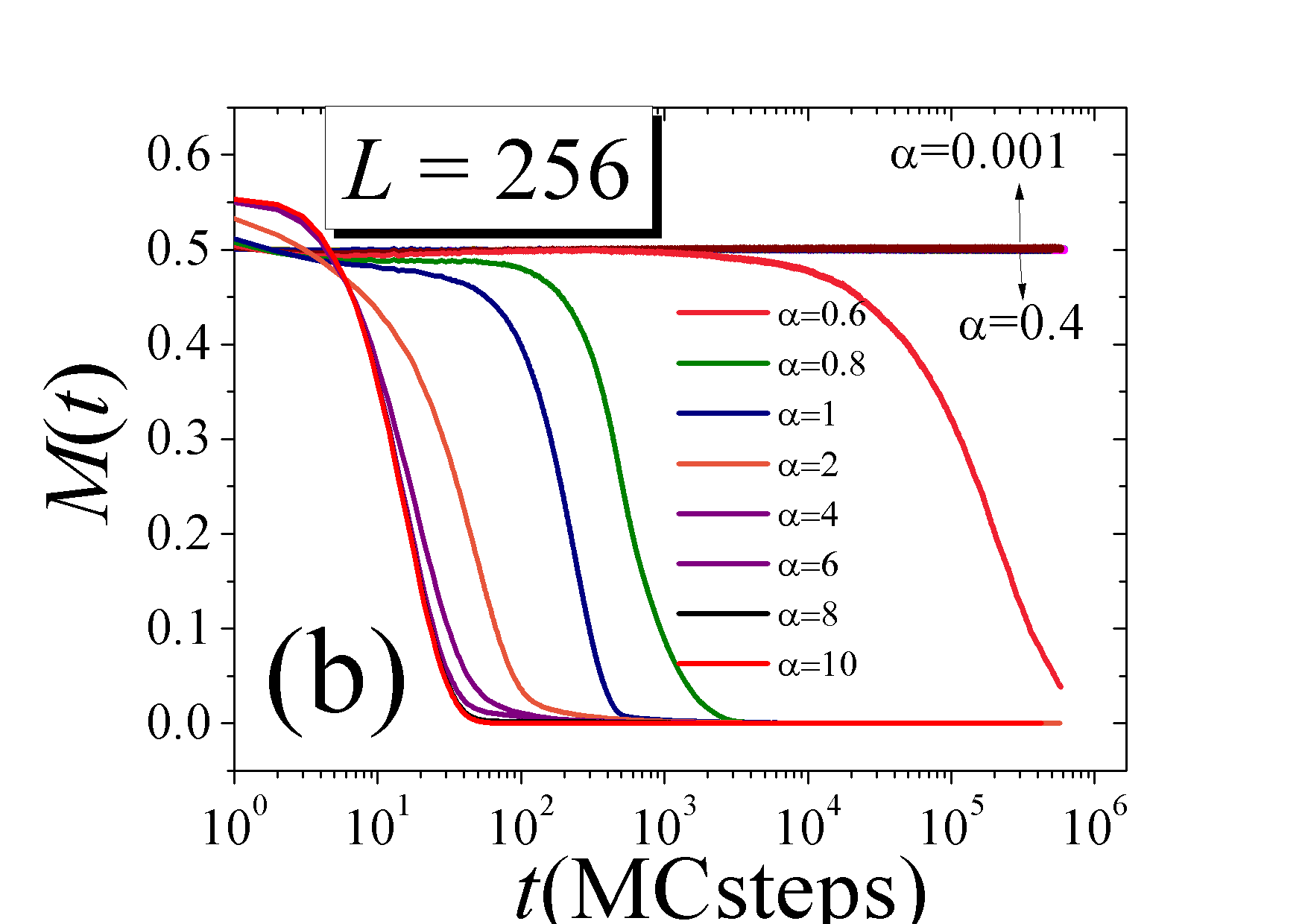} %
\includegraphics[width=1.0\columnwidth]{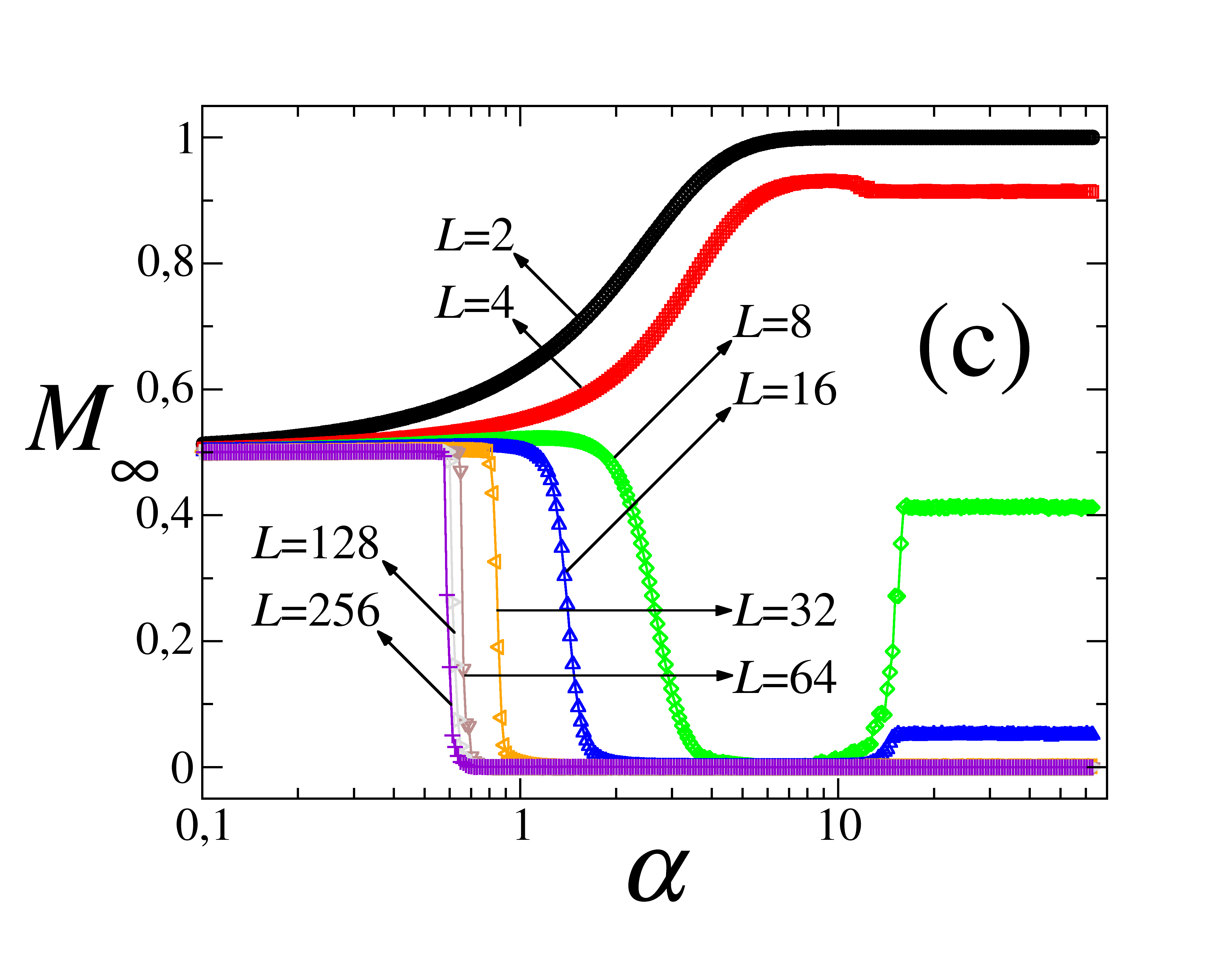}\includegraphics[width=1.0%
\columnwidth]{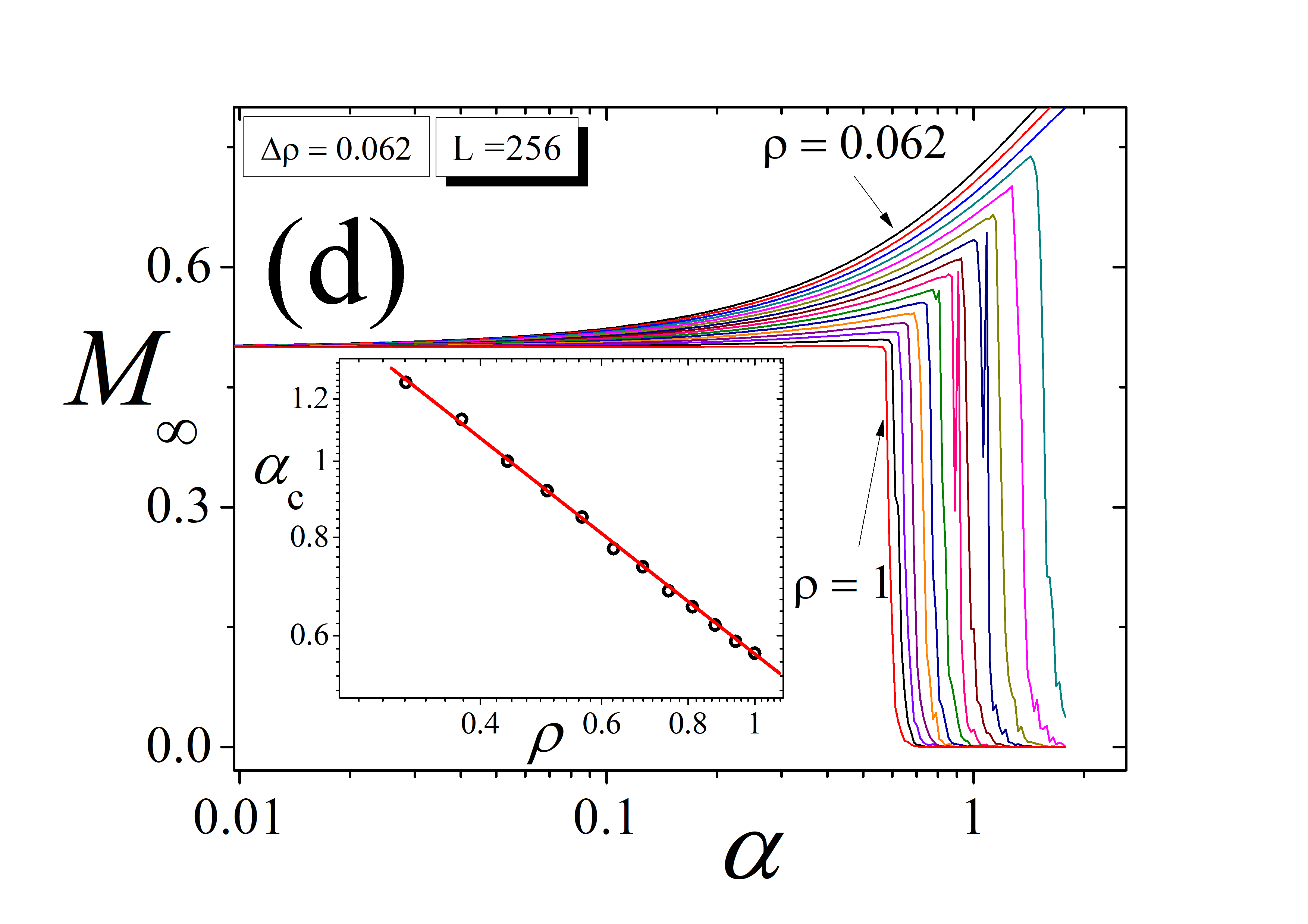}
\end{center}
\caption{(a) Time evolution of mobility for $L=8$. (b) Time evolution of
mobility for $L=256$. (c) Finite size scaling of stationary mobility as
function of $\protect\alpha $. (d) Stationary mobility as function of $%
\protect\alpha $ for fixed densities. An abrupt transition from mobile state
to clogging state for $\protect\alpha =\protect\alpha _{c}$, can be observed
which depends on density. The inset plot (in log-log scale) shows the
dependence of $\protect\alpha _{c}$ as function of $\protect\rho $. }
\label{Fig:mobility}
\end{figure*}

We performed simulations for several values of systems $L$. Fig. \ref%
{Fig:mobility} (a) and (b) show respectively the time evolution of mobility
for $L=8$ and $L=256$. Here we average the mobility over considerable number
of runs: $N_{run}=L^{-1}10^{6}$ runs. We can observe that plots are really
different. So it is interesting to check the stationary mobility for
different size systems as function of $\alpha $ which is observed in \ref%
{Fig:mobility} (c). We can check that system is deeply sensitive on the size
system, but for $L\geq 128$, no numerical differences were observed and Fig. %
\ref{Fig:mobility} (d) shows the results for $L=256$ considering different
densities from $\rho =0.062$ up to 1. Coming back to Fig. \ref{Fig:mobility}
(c), it is important to mention an anomalous recovering of mobility for
large values of $\alpha $, but such distortion only occurs for really small
systems, which is not relevant.   

These results show an abrupt transition between a mobile phase ($m_{\infty
}>0$) to a clogging phase ($m_{\infty }=0$). This transition is preceded by
an initial slip of mobility. This occurs because when the interaction of the
environment with objects decreases, i.e., $\alpha $ enlarges, the objects
initially gain mobility presenting an initial slip of the mobility, given
their high momenta. But as $\alpha $ enlarges even more, the interaction
among the objects really increases, until it finally destroys the mobility.
In this case the motion is random only when cell occupation assumes exactly
the value $\sigma _{\max }$. The high momenta of the objects that ignore the
environment is an important constraint to make the system reaches the
clogging situation due to the strong interaction effects among objects. It
is interesting to observe that the abrupt transition to a clogging phase
occurs with a high peak of density immediately followed by a large number of
smaller peaks of bottlenecks, as suggested by Figs. \ref%
{Fig:Different_frames_for_jamming} and \ref{Fig:Vicinity_of_transition}.

This analysis concerned $\sigma _{\max }=1$. So the question is, we should
observe anomalous effects for $\sigma _{\max }>1$, which in our systems
means to consider small objects or simply, more \textquotedblleft
particles\textquotedblright\ occupying the same \textquotedblleft
orbital\textquotedblright ? In this case we can observe a clogging
transition for a $\alpha _{c}^{(1)}$ and a recovering of mobility of the
system for a $\alpha _{c}^{(2)}>$ $\alpha _{c}^{(1)}$? Yes, it ocurrs. So we
analyze simulations of mobility considering now $\sigma _{\max }>1$. In this
case it is important to make a distinction, density and occupation. We
define as density $\rho =\frac{N}{L}$ where $N$ is the number of particles
and $L$ the system size, or simply the number of cells. Occupation is a
different concept, which here is defined as $o=\frac{N}{\sigma _{\max }L}$.
Thus we prepared two experiments, so that in one of them, we change $\sigma
_{\max }$ keeping the density constant, and in another simulation, we change 
$\sigma _{\max }$ keeping constant the occupation. And two surprising
results are observed.

\begin{figure}[tbh]
\begin{center}
\includegraphics[width=1.0\columnwidth]{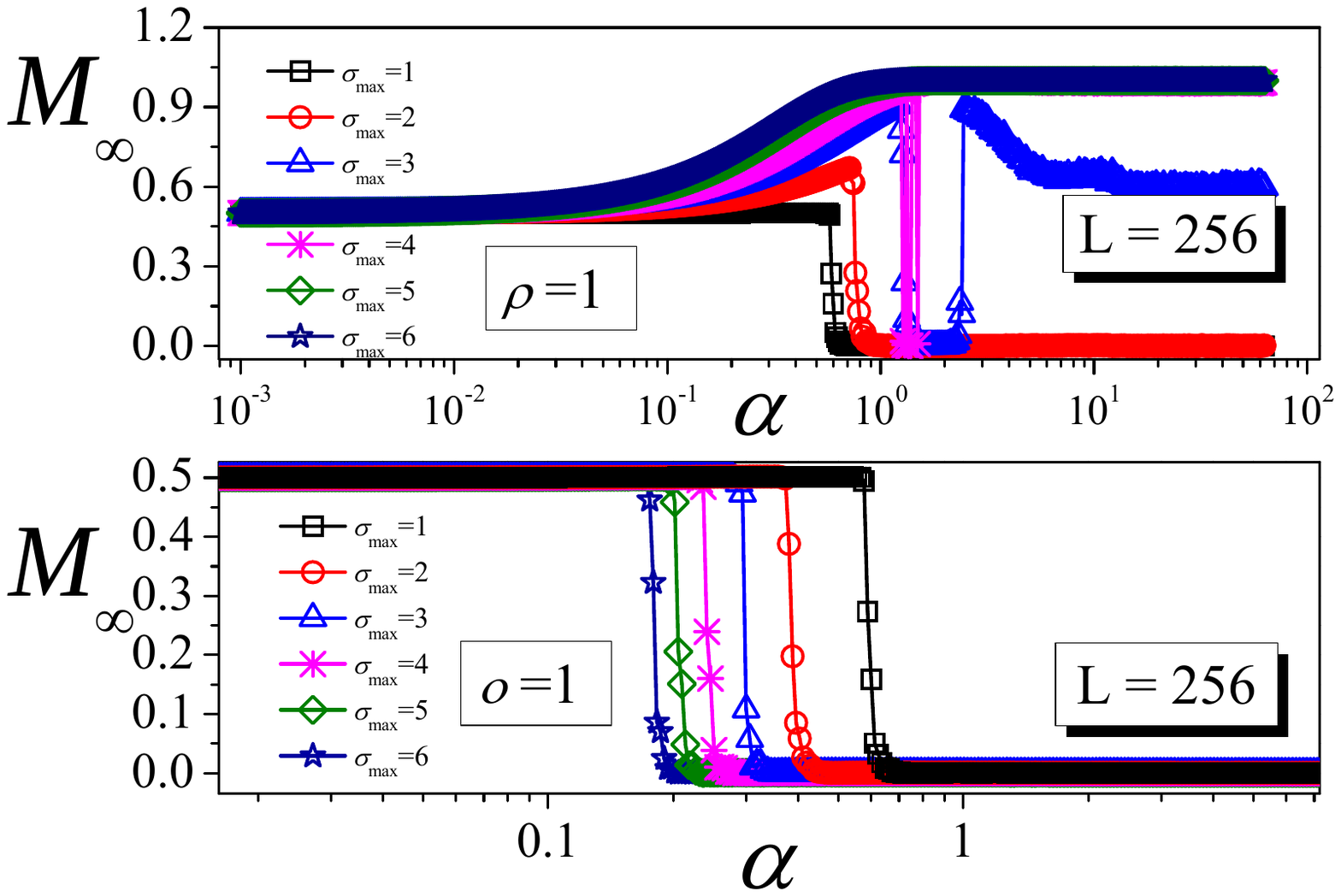}
\end{center}
\caption{Stationary mobility in two situations: $\protect\rho =1$ and $o=1$,
for different values of maximal ocupation: $\protect\sigma _{\max
}=1,2,...,6 $. }
\label{Fig:different_sigmas}
\end{figure}

Fig. \ref{Fig:different_sigmas} shows two distinct situations where we
variate $\sigma _{\max }$. First keeping the density constant at $\rho =1$
and in a second case, keeping the occupation constant at $o=1$. In the first
case we can observe a recovering of mobility for $\sigma _{\max }=3$ and $%
\sigma _{\max }=4$ and for $\sigma _{\max }>4$ the system does not present a
formation of condensates for no value of $\alpha $, i.e., we can transit
from a situation of objects interacting randomly with environment to
situation where the objects strongly interact among them without the
influence of the environment, and no bottleneck is observed since the object
size in relation to cell size allows such situation. However, in the
anomalous cases $\sigma _{\max }=3$ and $\sigma _{\max }=4$, the clogging
occurs as in the case $\sigma _{\max }=1$, but the mobility is recovered
(anomalously) for higher values of $\alpha $, given that the absence of
randomness from the environment combined with intermediate relation between
the object size and cell size. Here, differently from anomalies for very small 
systems, we believe that such point could also be due to an
artefact of the synchronous MC simulations, which should be fixed by
performing asynchronous MC simulations which deserves part of our future
attention. 

On the other hand, by keeping $o=1$, we did not wait a change in the
critical value $\alpha _{c}$, since we enlarge $\sigma _{\max }$ as well as
we enlarge the number of objects to maintain $o$ constant. This non-linear
response is characteristic of the Fermi-Dirac distribution of the cells for
the transition probability. Such effects deserve more future investigation.

\section{Summaries, conclusions, and discussions}

\label{Sec:Conclusions}

In this work we show a different model that works with a parameter $\alpha $%
\ that controls the randomness of the system by changing how the objects
interact with the environment and among themselves. We observe a transition
between a mobile phase and clogged phase in a $\alpha _{c}$\ that depends on
occupation of objects.

It is important to mention that such phenomena have some similarities with
analogous models. For example, Helbing, Farkas, and Vicsek \cite{Helbing2000}
considering a simple model of particles driven in opposite directions and
interacting via a repulsive potential, have found a transition to a
crystallized state from a fluid state by increasing the amount of
fluctuations of the system. The existence of condensates here observed as
function of $\alpha $, also was analyzed in other interesting and beautiful
way by Majundar, Evans, and Zia \cite{Majundar2005} considering the shape of
equilibrium mass distribution that change as the global mass density change.
However the authors have studied such model only considering a single
species which suggests that such model can be changed to cover our results.

By concluding, we believe that our model deserves more future explorations
to better understand the mobile-clogging transition in such systems of
counterflowing stream of particles via both analytical and computational
methods.

\textbf{Acknowledgements}

R. da Silva and E. V. Stock were financially supported by CNPq under grant numbers:
311236/2018-9, 424052/2018-0, and 154822/2016-7. This work was partly developed using 
the resources of Cluster Ada, IF-UFRGS. Finally, the authors are very grateful
to Prof. Sandra Prado for the helpful suggestions to this manuscript.

\end{document}